\documentclass[a4paper]{toledoStyle}
\usepackage{epsfig}

\begin{document}

\title{Simulating galaxy surveys with FIRST (PACS $\&$ SPIRE)}

\author{A.-L. Melchior\inst{1} \and F. Combes\inst{1} \and
B. Guiderdoni\inst{2} \and S. Hatton\inst{2}}

\institute{
  DEMIRM, Observatoire de Paris, 61,avenue de
l'Observatoire, 75014 Paris, France 
\and 
  IAP, 98bis, boulevard Arago, 75014 Paris, France}

\maketitle 

\begin{abstract}

The next generation of submillimetre/millimetre instruments will
provide us with a deeper insight into the mechanisms that rule galaxy
formation. As the brightest starbursts are thought to be heavily
obscured at optical wavelengths, the opening of this new window will
complement the present observations, and enable a detailed
investigation of the hierarchical merging of galaxies at remote
epochs. In this context, we will present preliminary results of
simulations currently developed for deep galaxy surveys with FIRST.\\
$N$--body simulations of galaxy mergers are being developed to produce
realistic morphologies and star formation histories. A Schmidt law is
used to account for the evolution of the stellar and gas content of
each particle. These simulations will be gathered in a library,
including the temporal information with a timestep of 10 Myr.  We then
build consistent spectra accounting for this star formation history at
each resolution element of the simulations. These morphologies will be
included in the framework of the {\sc GalICS} hybrid model of
hierarchical galaxy formation, which reproduces the main observational
constraints.  We intend to present synthetic maps with the
characteristics of the FIRST PACS/SPIRE instruments and discuss the
optimal strategy for deep surveys complemented with the ground-based
ALMA project.

\end{abstract}

\section{Introduction}
The understanding of galaxy formation and evolution is a rapidly
evolving field. The infrared/submillimetre sources detected with
SCUBA/JCMT are compatible with the IR Background detected with
COBE. The dusty galaxies, whose identification (redshift, morphology)
is so difficult (e.g. Downes et al. 1999), probably escape current
optical surveys. The new generation of instruments, which will become
available in the coming years, will provide us with an unprecedented
multi-wavelength view of these galaxies. From this perspective, we
have undertaken a comprehensive simulation work that will be used to
reproduce current observations and to predict forthcoming ones.

In this paper, we present the main framework of this study as well as
preliminary applications to FIRST. In Section \ref{simplemodel}, we
present a simple analytic model of starburst spectra and apply it to
known starburst galaxies. In Section \ref{N-body}, the simulations of
galaxy mergers being developed are described together with preliminary
applications. In Section \ref{GALICS}, we conclude with a short
overview of the general framework of the {\sc GalICS} hybrid model of
hierarchical galaxy formation, within which the previous work will be
incorporated.
\begin{figure}[ht]
  \begin{center}
    \epsfig{file=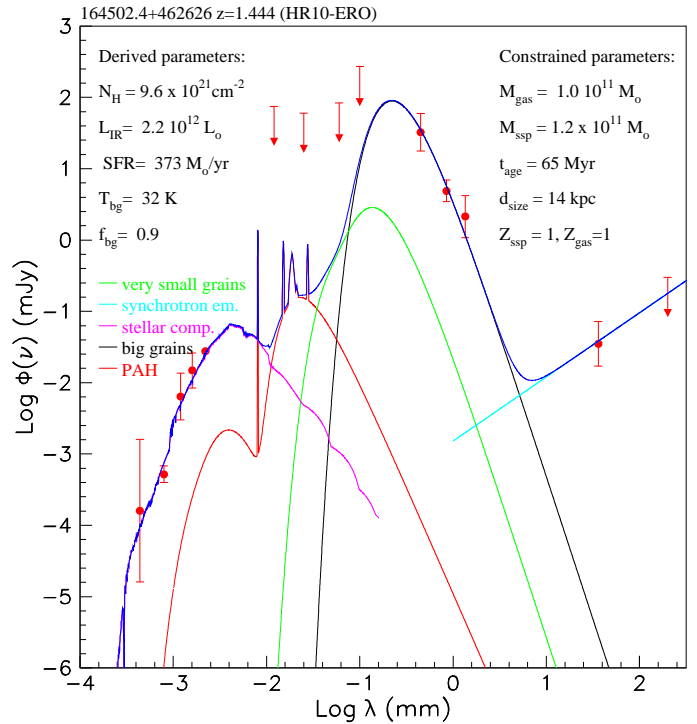, width=0.5\textwidth}
  \end{center}
\caption{Best model for the spectra of the ERO J164502+4626.4
(HR10). The measurements  have been published by
\protect \cite*{melchiora:dey1999}.}  
\label{melchiora_fig:fig1}
\end{figure}

\section{A simple model for galaxy spectra}
\label{simplemodel}
We have built an analytic tool to produce spectra of starbursts, that
will be used to produce images of galaxy mergers based on N-body/SPH
simulations. Each of these spectra will be associated with each
resolution element, and it is essential to have a restricted number of
parameters.  This minimal treatment is intended to model high-z
galaxies and to get a deeper insight into the process of star
formation.

\begin{figure}[ht]
  \begin{center}
    \epsfig{file=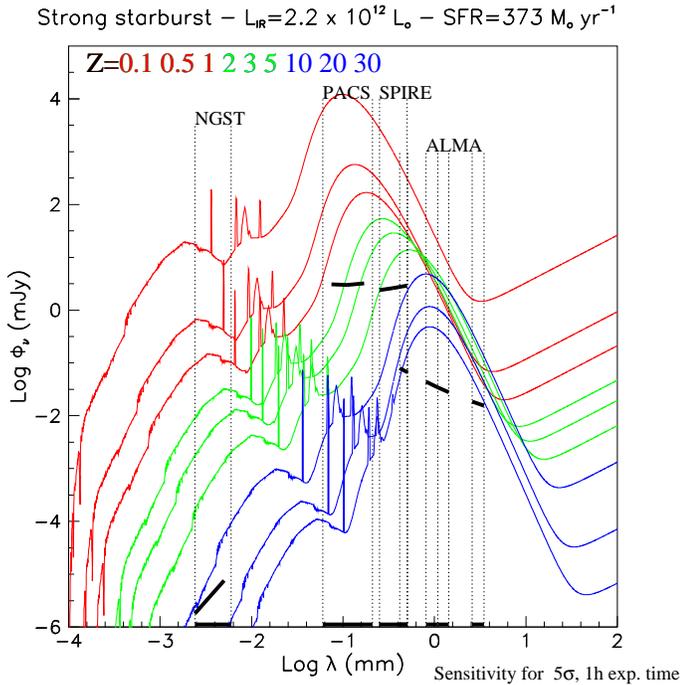, width=0.5\textwidth}
  \end{center}
\caption{A typical strong starburst galaxy (HR10) spectra is
extrapolated for different redshifts. The thresholds of PACS, SPIRE as
well as NGST and ALMA are presented. Such galaxies could potentially
be observed up to a redshift of 5.}
\label{melchiora_fig:fig2}
\end{figure}

\subsection{Assumptions}
First of all, we gathered a library of Simple Stellar Population (SSP)
starting from PEGASE.2 (Fioc $\&$ Rocca-Volmerange (1997)) for
different ages (1~Myr-20~Gyr) and different metallicities ($Z/Z_\odot$=
0.005, 0.02, 0.2, 0.4, 1., 2.5, 5). This library contains the sole
stellar component.  We then add the extinction based on local
star-forming galaxies computed by Calzetti et al. (2000) depending on
$N_H$ and the gas metallicity. The energetic balance allows us to
compute the luminosity $L_{IR}$, re-emitted in the infrared by the
dust. In a similar fashion as Guiderdoni et al. (1998) and Devriendt
et al. (1999) , we re-distribute this energy between 3 components
(namely big grains, very small grains and PAH), following D\'esert et
al. (1990). Last, the synchrotron emission is also included with a
$\nu^\beta$ law ($\beta=-0.7$) following Condon et al. (1991).

\subsection{Free  parameters}
This minimal modelling already has 6 free parameters: 
\begin{itemize}
\item $M_{ssp}$ : the mass of stars
\item $Z_{ssp}$ : the metallicity of the stars
\item $M_{gas}$ : the mass of gas
\item $Z_{gas}$ : the metallicity of the gas
\item $t_{ssp}$ : the age of the starburst
\item $d_{size}$ : the size of the galaxy or starburst encompassing the
previous gas and stars
\end{itemize}

\begin{figure}[ht]
  \begin{center}
    \epsfig{file=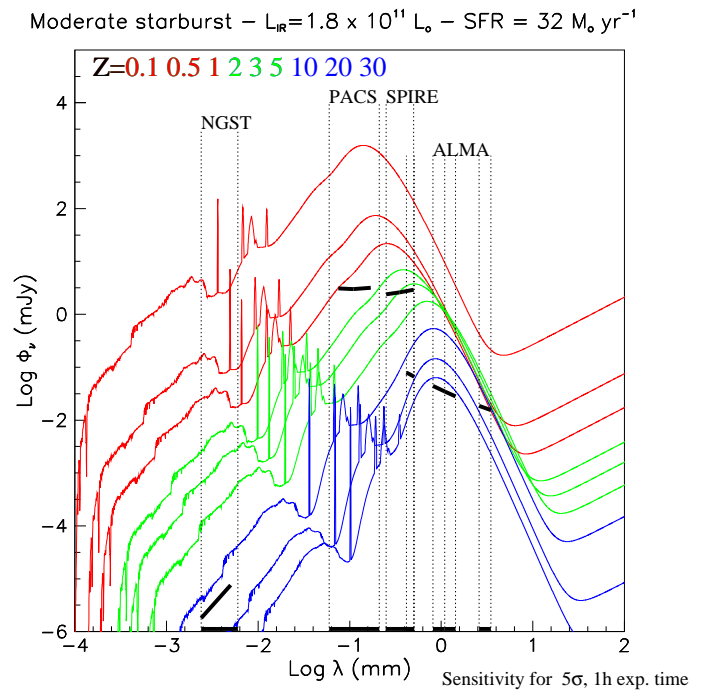, width=0.5\textwidth}
  \end{center}
\caption{Same as Fig. \protect\ref{melchiora_fig:fig2} but for a
moderate starburst. The characteristics are taken from NGC~6090 (at
$z=0.03$) as published by
\protect\cite*{melchiora:Calzetti2000}. Galaxies up to a redshift of 2
can then be studied.}
\label{melchiora_fig:fig3}
\end{figure}
\subsection{Modelling known galaxy spectra}
In order to test the validity of this simple model, we fit known
spectra of starburst galaxies. Given the small number of available
measurements, we deliberately keep the metallicities to solar values.
Firstly, we fit the optical part ($<2\mu$m) in order to adjust
$M_{ssp}$, $M_{gas}$ and $t_{ssp}$. Secondly, the
infrared/millimetre/radio measurements are used to define $d_{size}$.

The first example corresponds to a strong starburst with
SFR$\sim$370$M_\odot$yr$^{-1}$ (see Fig. \ref{melchiora_fig:fig1} and
\ref{melchiora_fig:fig2}) and the second one is a more moderate one with
SFR$\sim$30$M_\odot$yr$^{-1}$ (see
Fig. \ref{melchiora_fig:fig3}). Although real galaxies are probably
composed of old stellar populations and young starbursts, this simple
model based on starburst assumptions is already a good fit.

\subsection{Application to FIRST}
We use the spectra of Fig. \ref{melchiora_fig:fig2} and
\ref{melchiora_fig:fig3}  to study the sensitivity of PACS
and SPIRE to high-z galaxies.  These figures show starburst spectra at
different redshifts as presented by Combes et al. (1999) together with
the sensitivity of FIRST, ALMA and NGST. The wavelength coverage of
PACS and SPIRE will be ideal to sample the maximum of the spectral
luminosity distribution of starburst galaxies. These instruments will
be unique to understand the luminosity budget and properties of the
galaxies representative of the peak of the Madau curve (Madau et al.,
1998), as well as the excitation conditions, which determine the
infrared/millimeter flux. Indeed, as shown in Figures
\ref{melchiora_fig:fig2} and \ref{melchiora_fig:fig3}, a galaxy at
redshift 5, detected in the optical by the NGST, might escape
detection with FIRST if the starburst is not strong enough. In
parallel, ALMA will be able to detect starburst galaxies up to $z=30$
if they exist, though it will not be able to sample the peak below
$z=5$.
\begin{figure*}[ht]
  \begin{center}
    \epsfig{file=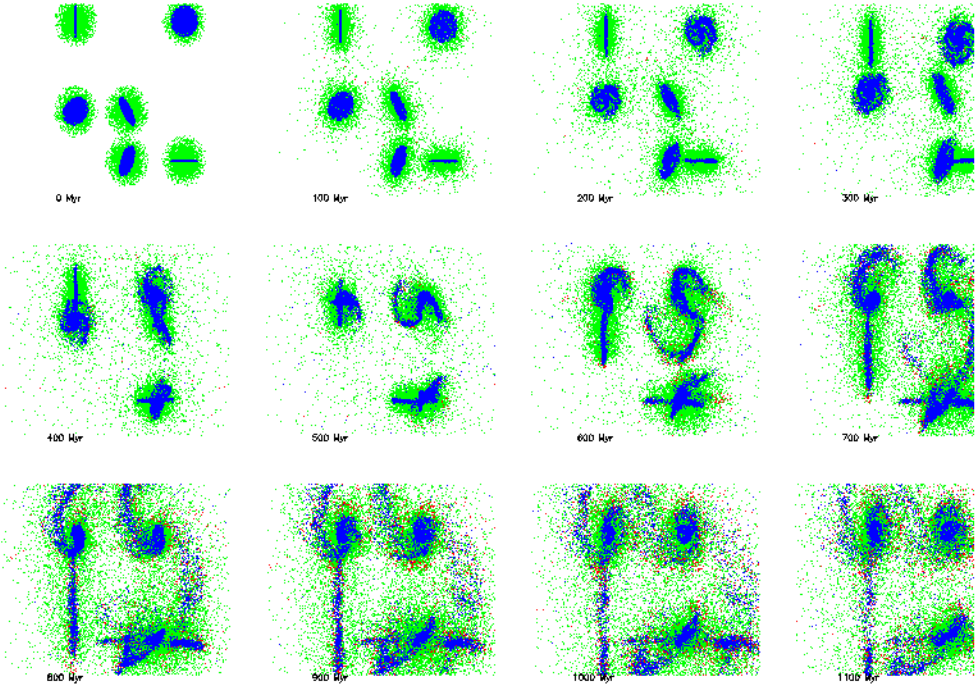, angle=-90, width=0.7\textwidth}
  \end{center}
\caption{Sequence of a simulated merger, showing the evolution with a
timestep of 100~Myr. Each panel presents a pair of galaxies in
interaction with 3 different orientations, as follows: upper
right=plane (XY), upper left=plane (ZY), bottom right=plane (XZ).  The
green points correspond to the dark matter, the blue ones to the gas
and the red ones to the stars.}
\label{melchiora_fig:fig4}
\end{figure*}

\section{N-body simulations of galaxy mergers}
\label{N-body}
\subsection{Principle}
We use these analytic starburst spectra to produce realistic maps of
galaxy mergers, based on N-body simulations. We start from one N-body
+ hydrodynamics simulation (tree-code + SPH) of a galaxy encounter
stored for different epochs. It is composed of 8\,000 dissipative
particles (hybrid: star or gas), 8\,000 stellar particles and 8\,000
particles for dark matter. We account for star formation with a
Schmidt law ($\rho_{star}
\propto \rho_{gas}^{1.5}$) for the hybrid particles, which thus
encompass a variable fraction of gas and stars. In order to account
for supernovae, the instant recycling approximation for the gas
rejection into the interstellar medium is used with standard values
for the yield.

Figure \ref{melchiora_fig:fig4} displays preliminary simulations
without the star formation and metallicity enrichment prescription for
different epochs as well as different orientations. Although a single
simulation can already be used to simulate a large variety of
morphologies, we plan to generate several such simulations in order to
span a reasonable range of possible morphologies.

\subsection{Application}
\begin{figure}[ht]
  \begin{center}
    \epsfig{file=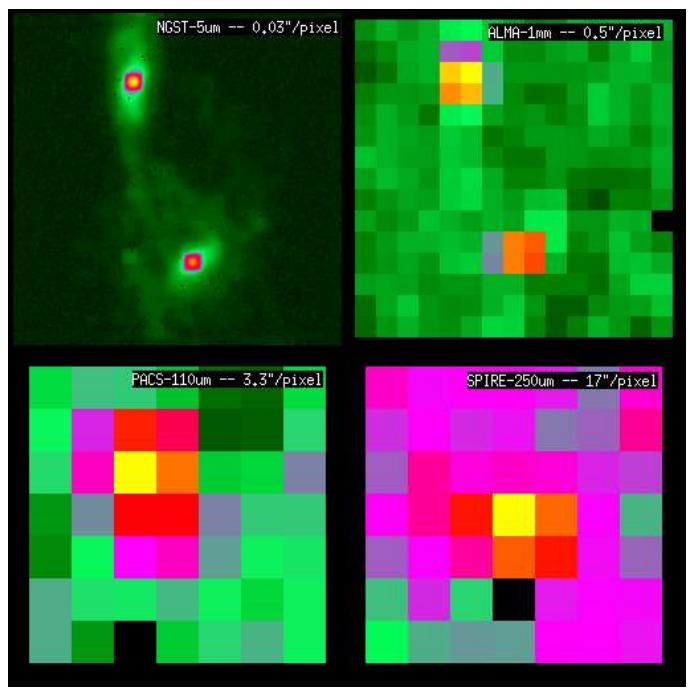, width=0.5\textwidth}
  \end{center}
\caption{Preliminary projection of a merger at $z=1$ ($L_{IR} =
2 \times 10^{12}L_\odot$) on a pixel grid with the characteristics of
FIRST/PACS $\&$ SPIRE, NGST and ALMA. These images correspond to 1h
integration time and account for the expected rms intrumental noise
for each intrument.}
\label{melchiora_fig:fig5}
\end{figure}

Figure \ref{melchiora_fig:fig5} presents a very preliminary
application of such a simulation for different epochs. Maps of such a
merger at $z=1$ are displayed as it could be observed by FIRST (PACS
$\&$ SPIRE), NGST or ALMA. Whereas the NGST will be able to study the
detailed morphology and possibly tidal arms, ALMA, after 1h of
integration time, will only be able to detect the multiplicity of the
source. The instruments of FIRST will sample the luminosity peak of
the spectral energy distribution but will not be able to distinguish
any signature of merging.

We currently extend this kind of simulation in a systematic way
(different morphologies) and from a cosmological perspective (within
{\sc GalICS} framework) in order investigate the current open
questions.  Firstly, the galaxies that will be observed in the optical
(e.g. with the NGST) will not be necessarily accessible to the IR/mm
instruments, especially if the star formation activity is
moderate. Secondly, the difficulties encountered to find optical
counterparts for SCUBA sources are hoped to be overcome with the next
generation of instruments. This simulation work will help to
understand the multi-wavelength dependence of these high-$z$ galaxies
and the process of galaxy formation.

\section{{\sc GalICS} model}
\label{GALICS}
The ``hybrid'' approach is a powerful tool to describe hierarchical
galaxy formation. The {\sc GalICS} model (for {\it Galaxies in
Cosmological Simulations}, Hatton {\it et al.}  2001, Devriendt {\it
et al.} 2001) uses a large cosmological $N$--body simulation ($256^3$
particles in a 150 Mpc box within a $\Lambda$CDM cosmology) achieved
on the CRAY--T3E of the IDRIS computation centre.  Halos with more
that 20 particles are identified by the usual friend--of--friend
algorithm, and a catalogue of haloes with a minimum halo mass of $1.7
\times 10^{11} M_\odot$ is built for each of the $\sim 100$ time
outputs. Merging history trees are built from these 100
catalogues. Finally, the evolution and merging history trees are built
for the galaxies.  The processes of dissipative gas cooling and
collapse, star formation, chemical evolution, stellar feedback are
modelled along lines which are detailed in the papers. The UV to mm
spectral energy distributions are computed from the {\sc stardust}
spectra (Devriendt {\it et al.} 1999).  After halo merging, the
satellite galaxies spiral down to the central galaxy with the
dynamical friction time scale. Satellite--satellite merging is also
taken into account. During the major mergers, a starburst light up and
radiates mostly in the IR/submm range, and a bulge eventually forms. A
new disk can form around this galaxy if it is at the cooling centre of
its halo. The morphologies of the galaxies are fixed according to the
$B$--band bulge to disk ratios. The final output at $z=0$ includes
about 70\,000 galaxies, and the mass resolution progagates as an
absolute magnitude resolution of $M_B=-18$.  Fake images of deep
fields can be achieved at any wavelength from the UV to the mm range
from the output catalogues.

So far, the code includes only very crude recipes on how galaxies
merge and how merging triggers a starburst event. As a first step, it
is planned to generate fake images of deep galaxy fields at optical
and IR/submm wavelengths from the catalogue of morphologies of
interacting and merging galaxies described in section 3, by choosing
the closest case to the initial conditions for the encounter fixed by
{\sc GalICS}. As a second step, the tree--code $+$ SPH hydrodynamical
simulations can be directly run from the galaxy merging trees with the
initial conditions of {\sc GalICS}.  Such a program will allow us to
assess the properties of mergers in a cosmological framework, and to
study how this properties evolve with redshift.  These fake fields
will be used to study the observational strategy for deep galaxy
surveys with {\it PACS} and {\it SPIRE}.

\end{document}